\documentclass{elsart}
\usepackage{natbib}
\usepackage{epsfig}
\begin{document}
\begin{frontmatter}
\title{High Resolution Hybrid  Pixel Sensors for the\\
$e^+e^-$ TESLA Linear Collider Vertex Tracker}
\author{M. Battaglia, R. Orava, K. Tammi, K. \"Osterberg}
\address{Department of Physics and Helsinki Institute of Physics,\\
         P.O. Box 9, FIN--00014 University of Helsinki, Finland}
\author{W. Kucewicz}
\address{Department of Electronics, University of Mining and Metallurgy,\\
al. Mickiewicza 30, PL--30055 Krakow, Poland}
\author{A. Zalewska}
\address{High Energy Physics Laboratory, Institute of Nuclear Physics,\\
ul. Kawiory 26a, PL--30055 Krakow, Poland}
\author{M. Caccia, R. Campagnolo, C. Meroni}
\address{Dipartimento di Fisica, Universita' di Milano and I.N.F.N.,\\
                  Via Celoria 16, I--20133 Milano, Italy}
\author{P. Grabiec, B. Jaroszewicz, J. Marczewski}
\address{Institute of Electron Technology,\\
al. Lotnikov 32/46, PL--02468 Warszawa, Poland}
%============================================================================
%
%============================================================================
\begin{abstract}
In order to fully exploit the physics potential of a future high energy 
$e^+e^-$ linear collider, a Vertex Tracker, providing high resolution track
reconstruction, is required.
Hybrid Silicon pixel sensors are an attractive option, for the sensor
technology, due to their read-out 
speed and radiation hardness, favoured in the high rate environment of the 
TESLA $e^+e^-$ linear collider design, but have been so far limited by the 
achievable single point space resolution. In this paper, a conceptual design 
of the TESLA Vertex Tracker, based on a novel layout of hybrid pixel sensors 
with interleaved cells to improve their spatial resolution, is presented.
\end{abstract}
\end{frontmatter}
\section{The requirements of the linear collider physics programme}

The next generation of high energy $e^{+}e^{-}$ experiments,
following the LEP and SLC programs, will be at a linear collider, operating 
at centre-of-mass energies ranging from the $Z^0$ pole up to about 1~TeV. 
Expected to be commissioned by the end of the first decade of the new
millennium, the linear collider will complement the physics reach of the 
Tevatron and LHC hadron colliders in the study of the mechanism of 
electro-weak symmetry breaking and in the search for new physics beyond the 
Standard Model. Both precision measurements and particle searches set
stringent requirements on the efficiency and purity of the flavour 
identification of hadronic jets since final states including short-lived $b$ 
and $c$-quarks and $\tau$ leptons are expected to be the main signatures. High
accuracy in the reconstruction of the charged particle trajectories close to 
their production point is required in order to reconstruct the topologies of 
the secondary vertices in the decay chain of short-lived heavy 
flavour particles.

If a Higgs boson exists with mass below 150~GeV/c$^2$, as indicated by the 
fit to the present electro-weak data~\cite{Gross}, it will be essential to 
carry out precision measurements of its couplings to different fermion species
as a proof of the mass generation mechanism and to identify its Standard Model
or Supersymmetric nature~\cite{higgs1}. This can be achieved by accurate 
determinations of its decay rate to $b\bar{b}$, $c\bar{c}$, 
$\tau^{+}\tau^{-}$, $W^{+}W^{-}$ and gluon pairs to detect possible deviations
from the Standard Model predictions \cite{higgs2}. Since the rates for the 
Higgs decay modes into lighter fermions $h^0 \rightarrow c \bar c$, 
$\tau^+ \tau^-$ or into gluon pairs are expected to be only about 10\% or less 
of that for the dominant $h^0 \rightarrow b \bar b$ process, the extraction 
and measurement of the signals of these decay modes requires suppression of 
the $b \bar b$ contribution by a factor of twenty or better while preserving a 
good efficiency. 

The measurement of the top Yukawa coupling~\cite{TopYuk} as well as the 
top-quark mass measurement will require efficient $b$-tagging to reduce 
combinatorial background in the reconstruction of the six and eight jet final
states. 
If Supersymmetry is realized in nature, the study of its rich Higgs sector will
 also require an efficient identification of $b$-jets and $\tau$ leptons to 
isolate the signals for the decays of the heavier $A^0$, $H^0$ and $H^{\pm}$ 
bosons from the severe combinatorial backgrounds in the complex 
multi-jet hadronic final states. 
Due to the expected large $b$-jet multiplicity, highly 
efficient tagging is required to preserve a sizeable statistics of the 
signal events. Finally, both $b$ and $c$-tagging will be important in the study
of the quark scalar partners, while $\tau$ identification may be 
instrumental in isolating signals from Gauge Mediated Supersymmetry 
Breaking~\cite{GMSB}. 

\begin{figure}
\begin{center}
\epsfig{file=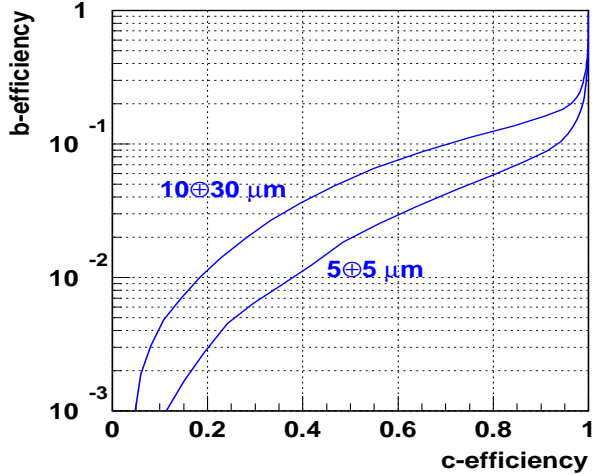,height=8.0cm,width=9.5cm}
\caption{The $b$-quark background efficiency as a function of the $c$-quark 
signal efficiency for a $c$-tag algorithm based on topological vertexing and 
impact parameters,
developed on the basis of the experience from the SLC and LEP experiments, 
\cite{higgs3}. The two curves show the expected performance for two different
assumptions on the track impact parameter resolution.}
\label{fig:ctag}
\end{center}
\end{figure}

A set of curves representative of the performance of a jet flavour tagging 
algorithm at the linear collider are given in Figure~\ref{fig:ctag} under
two different assumptions for the track impact parameter resolution. 
The requirement of efficient $c$ identification with a rejection
factor against $b$-jets of more than ten, highlights the need of track 
impact parameter resolution $\sigma_{I.P.}$ better than $10~\mu m \oplus 
\frac{30~\mu m~{\mathrm GeV/c}}{{\mathrm p}~\sin^{3/2}\theta}$ in both 
projections. It is important to minimise the multiple scattering contribution
to the impact parameter resolution. In fact, $b$/$c$
discrimination is obtained mostly by probing the difference in charged decay 
multiplicities and invariant masses of the $B$ and $D$ hadrons. This requires 
to identify the majority of charged decay products that have typical momenta of
a few GeV/c in multi-jet events. 
A precise determination of the track perigee parameters close to 
their production point also assists in the track reconstruction and improves
the momentum resolution. The addition of the Vertex Tracker space points with 
7~$\mu$m resolution to those from a TPC main tracker improves the estimated 
momentum resolution $\sigma_p/p^2$ from 1.5~$\times$~10$^{-4}$ (GeV/c)$^{-1}$
to 0.6~$\times$~10$^{-4}$ (GeV/c)$^{-1}$ for charged particles of large 
momentum.
%, thus improving the Higgs mass 
%measurement using the dilepton recoil mass distribution, in the process 
%$e^+e^- \rightarrow h^0 Z^0$; $Z^0 \rightarrow \ell^+ \ell^-$ with 
%($\ell$ = e, $\mu$).  
The design of the linear collider Vertex Tracker and the choice of the sensor 
technology are driven by these requirements, to be achieved within the 
constraints set by the accelerator induced backgrounds at the interaction
region and by the characteristics of the physics events.
In the next section these constraints are discussed with specific reference
to the TESLA linear collider project, while section~3 presents the 
conceptual design proposed for the Vertex Tracker.
The silicon pixel sensor technology, developed to overcome the 
hybrid pixel sensor limitations in terms of single point space resolution,
is discussed in details in section~4.

\section{The experimental conditions at the TESLA interaction region}

Several concepts have been developed for the acceleration, preservation 
of low emittance and final focusing of electron and positron beams at 
energies in excess of 200~GeV/beam~\cite{lcreview}. The TESLA 
project~\cite{CDR} has proposed to use superconducting accelerating 
structures, operating at L-band frequency that delivers very long beam pulses 
($\sim$ 900 $\mu$s) accelerating up to 4500 bunches per pulse. This scheme 
allows a large bunch spacing (190-340~ns) making it possible to resolve single 
bunch crossings (BX) and also to perform fast bunch-to-bunch feedback needed 
to stabilize the beam trajectory within a single pulse, thus preserving the 
nominal luminosity of 3-5~$\times 10^{34}$~cm$^{-2}$~s$^{-1}$. 
The large luminosity of each individual bunch-crossing (
2.2~$\times~10^{-3}$~nb$^{-1}$~BX$^{-1}$) and the large number of bunches in a 
single pulse imply a high rate of background events that need to be 
minimised by identifying the bunch corresponding to the physics event of
interest.

A primary source of background at the linear collider interaction region 
is the incoherent pair production in the electromagnetic interactions of 
the colliding beams. These particles are confined by the solenoidal magnetic 
field, spiralling in an envelope defined by the field strength, $B$, and
their intrinsic transverse momentum acquired at production and the 
subsequent deflection in the electric field of the opposite beam. The pairs 
represent an irreducible source of spurious particle hits, which interferes 
with the reconstruction of the particles from the physics processes of 
interest.
The maximum radius and longitudinal position of crossing of the envelope of 
deflected pairs defines both the inward bound for the first sensor layer and 
its maximum length. At 1.2~cm radial distance from the TESLA colliding 
beams, with a magnetic field $B$ = 3~T, the hit pair density is expected to be
$\simeq$ 0.2 hit mm$^{-2}$ BX$^{-1}$ with $\pm~5$~cm available in the 
longitudinal coordinate to fit the first sensitive layer of the detector. 
An additional background source that needs to be taken into account in the 
choice of the sensor technology is the flux of neutrons photo-produced at
the dump of electrons from pairs and radiative Bhabha scattering and of 
beamstrahlung photons. The computation of this neutron flux at the interaction
region relies on the modelling of their production and transport in the 
accelerator tunnel and in the detector and is subject to significant 
uncertainties. Estimated fluxes are of the order of a few 
10$^9$ $n$~(1~MeV)~cm$^{-2}$~year$^{-1}$~\cite{Tesch,Ye}, where the 
anticipated neutron flux has been normalised in terms of equivalent
1~MeV neutrons assuming NIEL scaling.
Finally the large $e^+e^- \rightarrow \gamma \gamma$ cross-section requires
single bunch identification and high resolution on the longitudinal position
of production of forward charged particles in order to suppress to a few \% 
the rate of two photon background overlap with physics events.

In addition to these background sources, the occupancy from the charged 
particles 
inside dense jets in multi-parton hadronic final states has to be considered. 
A study of $t \bar t$ and $h^0 Z^0$ events at $\sqrt{s}$ = 500 GeV showed 
that at 3.0~cm from the interaction point, about 20 \% (10 \%) of the 
particles in the jet have at least one additional hit from another particle 
of the same jet within a 150 $\mu m$ distance in the $R-\Phi$ ($R-z$) plane, 
which corresponds to the typical two-track separation capability of a 
microstrip detector. Therefore,  sensors with small sensitive cells have to 
be used in order to avoid a large number of merged hits and ambiguities in the
pattern recognition.

In summary, the linear collider Vertex Tracker must be able to provide
a track impact parameter resolution better than $10~\mu m \oplus 
\frac{30~\mu m~{\mathrm GeV/c}}{{\mathrm p}~\sin^{3/2}\theta}$ in both 
the $R-\Phi$ and $R-z$ projections for jet flavour identification, identify 
single bunch crossings separated by about 200~ns to reduce pair and 
$\gamma \gamma$ backgrounds and have sensitive cells of 150 $\times$ 
150~$\mu$m$^2$ or less to keep the occupancy from pairs and hadronic jets
below 1\%. There are two types of such silicon sensors, already used at
collider experiments, that have the potential to satisfy these 
specifications in terms of sensitive cell size: the Charged Coupled Devices 
(CCD) and hybrid pixels sensors. The CCD sensors have successfully been used 
for the SLD Vertex Detector at the SLC collider at SLAC~\cite{vxd3}
while hybrid pixel sensors, pioneered by the WA-97 experiment at 
CERN~\cite{WA97}, have been adopted at LEP in the upgraded DELPHI Silicon 
Tracker~\cite{SiTracker}. These have been further developed for the 
ALICE~\cite{alice}, ATLAS~\cite{atlas} and CMS~\cite{cms} experiments to meet 
the experimental conditions of the LHC collider. 
CCD detectors have been already proposed for the linear collider Vertex 
Tracker~\cite{ccdlc}. While CCD's have ideal characteristics 
in terms of spatial resolution and detector thickness, they presently lack 
the required read-out speed necessary to cope with the TESLA bunch timing and
are possibly sensitive to neutron damage at fluxes of the order of that 
expected at the linear collider. An intense R\&D program is presently
underway to overcome these limitations~\cite{CCDnew}.
The use of hybrid pixel sensors for the linear collider Vertex Tracker was
also proposed a few years ago and a first conceptual design of the detector
was defined~\cite{ABC}. Compared to the CCD's, hybrid pixel sensors 
have the advantage of allowing fast time stamping and sparse data scan 
read-out, thereby reducing the occupancies due to backgrounds, and of being 
tolerant to neutron fluxes well beyond those expected at the linear collider. 
Both these characteristics have been demonstrated for their application in the 
LHC experiments. On the other hand, there are areas of R\&D that are 
specific to the linear collider, namely the improvement of the pixel sensor 
spatial resolution and the reduction of its total thickness.

\section{The TESLA Vertex Detector design}

\begin{figure}[h!]
\begin{center}
\epsfig{file=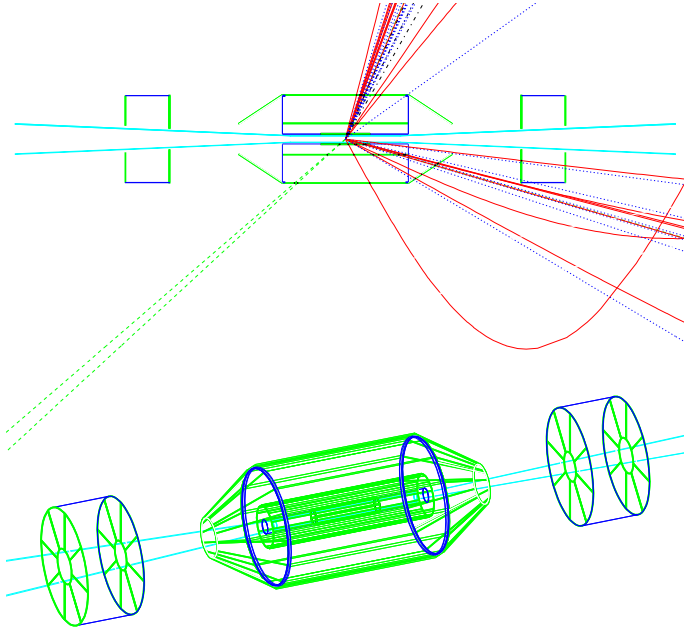,height=9.0cm}
\end{center}
\caption{The proposed layout of the Vertex Tracker with a simulated
$e^+e^- \rightarrow h^0 Z^0 \rightarrow b \bar b \mu^+ \mu^-$ event
overlayed. The three layered barrel
section is complemented by a crown and two disks of detectors to ensure 
accurate tracking in the forward region.}
\label{fig:layout} 
\end{figure}

\begin{figure}[h!]
\begin{center}
\epsfig{file=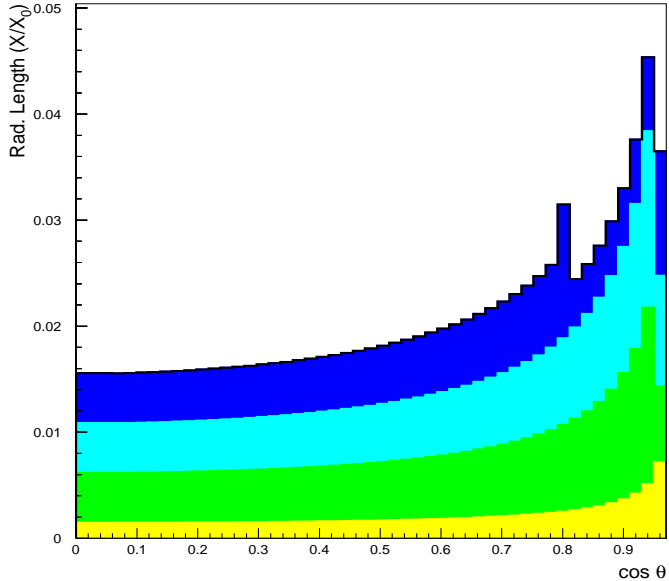,bbllx=25,bblly=160,bburx=535,bbury=660,width=9.0cm,
height=8.0cm}
\end{center}
\caption{The material budget given in units of radiation length as a function 
of $|\cos \theta |$. The contributions of a 0.5~mm Be beam-pipe and of the 
three detector layers with cables and support structure have been included.}
\label{fig:budget} 
\end{figure}

The proposed layout of the TESLA Vertex Detector \cite{ABC} based on hybrid
pixel sensors is shown in fig.~\ref{fig:layout} and consists of
a three-layer cylindrical detector surrounding the beam-pipe complemented
by forward crowns and disks extending the polar acceptance to small angles. 
This geometry follows solutions adopted for the DELPHI Silicon Tracker. 
The first detector layer, closest to the interaction
region, is located at a radius of 1.2~cm and has a length of 10~cm. 
The two additional barrel layers are located at 3.5~cm and 10~cm respectively
and have a polar acceptance down to $|\cos \theta| = 0.82$. At lower angles, 
additional space points are obtained by extending the barrel section by a 
forward crown and two disks of detectors providing three hits down 
to $|\cos \theta| = 0.995$. The transition from the barrel cylindrical to
the forward conical and planar geometries optimises the angle of incidence of 
the particles onto the detector modules in terms of achievable single point 
resolution and multiple scattering. This tracker can be assembled by two
independent half-shells, allowing its installation and removal with the 
beam-pipe in place. Overlaps of neighbouring detector modules will provide an
useful mean of veryfing the relative detector alignment using particle
tracks from dedicated calibration runs taken at $Z^0$ centre-of-mass energy. 
The geometry optimisation and the study of
the physics performances of the Tracker design has been performed with a 
GEANT based simulation of the detector, accounting for benchmark 
physics processes with enhanced forward production cross-section, such as 
$e^+e^- \rightarrow H^0 \nu \bar \nu$, pair and $\gamma \gamma$ backgrounds, 
local pattern recognition and detector inefficiencies. The impact parameter
resolution has been obtained by a Kalman filter track fit to the associated 
hits in the Vertex Tracker and in the TPC.
\begin{figure}[h!]
\begin{center}
\epsfig{file=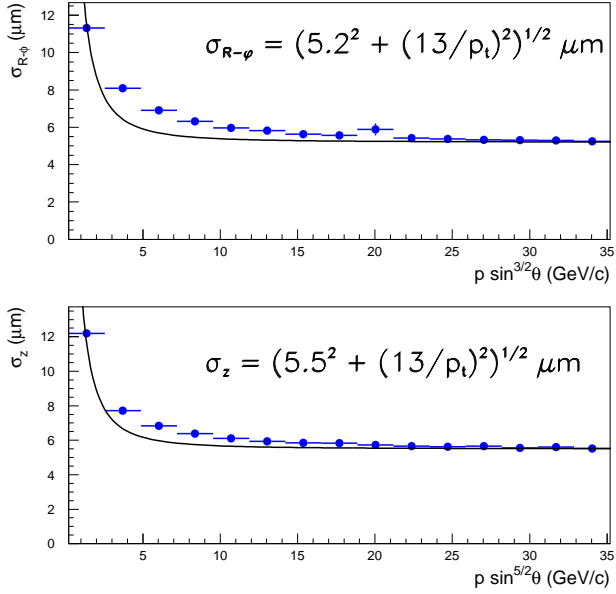,width=9.0cm,height=9.0cm}
\end{center}
\caption{The expected impact parameter resolution $\sigma_{I.P.}$ in the 
$R-\phi$ (upper plot) and $R-z$ (lower plot) projections as a function of the 
particle momentum for the proposed Vertex Tracker assuming a single point
resolution of 7~$\mu$m corresponding to that obtained for microstrip sensors
with 25~$\mu$m strip pitch and 50~$\mu$m read-out pitch. The curve overlayed
to each plot shows the result of a fit to the function
$\sigma_1 \oplus \frac{\sigma_2}{{\mathrm p}~\sin^{3/2 (5/2)}\theta}$. The 
departure of the points from the predictions of these parametrisations in the 
region from 2.0~GeV/c to
10~GeV/c is due to the effect of multiple scattering between the Vertex
Tracker and the TPC.}
\label{fig:ip} 
\end{figure}
In order to achieve a resolution for the impact parameter reconstruction 
better than 7~$\mu$m for tracks at large momenta, a detector space point 
accuracy of better than 10~$\mu$m must be obtained. The requirement on the 
multiple scattering contribution to the track extrapolation resolution, lower 
than 30~$\mu$m/$p_t$, and the need to minimise the amount of material in front
of the calorimeters and to ensure the optimal track matching with the main 
tracking system, set a constraint on the material budget of the Vertex 
Tracker to less than 3~\% of a radiation length ($X_0$). 
These requirements can be fulfilled by adopting 200~$\mu$m thick detectors and 
back-thinning of the read-out chip to 50~$\mu$m, corresponding to 0.3~\%~$X_0$
of a radiation length, and a light support structure. The present concept for
the mechanical structure envisages the use of diamond-coated carbon fiber
detector support layers acting also as heat pipes to extract the heat 
dissipated by the read-out electronics uniformly distributed over the whole
active surface of the detector.
Assuming a power dissipation of 60~$\mu$W/channel, the total heat flux is 
450~W, corresponding to 1500~W/m$^2$, for a read-out pitch of 
200~$\mu$m. Preliminary results from a finite element analysis show that pipes 
circulating liquid coolant must be placed every 5~cm along the longitudinal 
coordinate except for the innermost layer where they can
be placed only at the detector ends to minimise the amount of material.
Signals will be routed along the beam pipe and the end-cap disks to the 
repeater electronics installed between the Vertex Tracker and the forward
mask protecting the Vertex Tracker from direct and backscattered radiation
from the accelerator. The material budget for the proposed design is shown 
in Figure~\ref{fig:budget}.

\section{Hybrid pixel sensors with interleaved pixels}

The desired impact parameter resolution defined in section~1 
requires a single point precision in the Vertex Detector better than
${\rm 10~\mu m}$. This can be accomplished sampling the diffusion of the 
charge carriers generated along the particle path and 
assuming an analog read-out to 
interpolate the signals of neighbouring cells. In such a case, the expected
resolution is ${\rm \sigma = a_{cf} \frac{pitch}{S/N}}$, where 
${\rm a_{cf}} \approx 2$ is a centroid finding constant and S/N defines the
performance of the front-end electronics in terms of signal amplitude
normalised to the noise. Given that the charge diffusion is 
$\sim {\rm 8~\mu m}$ in ${\rm 300~\mu m}$ thick silicon, its efficient 
sampling and signal interpolation requires a pitch of not more than 25~$\mu$m.
This has been successfully proven to work in one-dimensional microstrip 
sensors \cite{Anna}. In pixel devices the ultimate read-out pitch is 
constrained by the front-end electronics, to be integrated in a cell matching 
the sensor pattern. At present, the most advanced read-out electronics have a 
minimum cell dimension of ${\rm 50 \times 300~\mu m^{2}}$ not suitable for an 
efficient charge sampling. The trend of the VLSI development and recent studies
\cite{Snow} on intrinsic radiation hardness of deep sub-micron CMOS technology 
certainly allows to envisage of a sizeable reduction in the cell dimensions 
on a linear collider timescale but sensor designs without such basic 
limitations are definitely worth being explored. 

\begin{figure}
\begin{center}
\epsfig{file=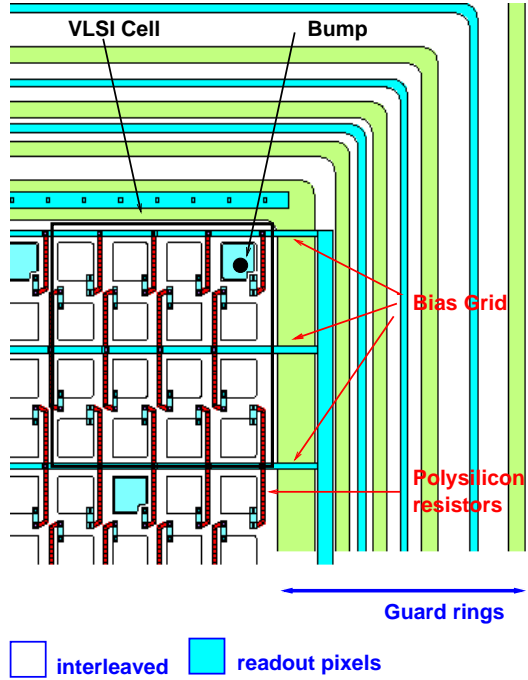,height=9.0cm}
\caption{Layout of the upper corner of pixel detector test structure, 
with 50~$\mu$m implant and 200~$\mu$m read-out pitch.}
\label{fig:corner}
\end{center}
\end{figure}

\begin{figure}[hb!]
\begin{center}
\epsfig{file=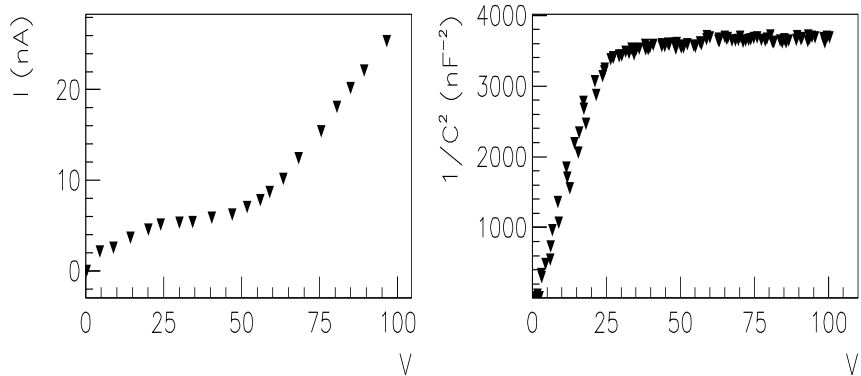,height=6.0cm,width=12.0cm}
\caption{I--V and 1/C$^{2}$--V curves for a good sensor structure}
\label{fig:ivcv}
\end{center}
\end{figure}

A possible way out is to exploit the capacitive coupling of neighbouring 
pixels and to have a read-out pitch n times larger than the implant 
pitch~\cite{Bonvicini}. The proposed sensor layout is shown in 
Figure~\ref{fig:corner} for n=4. In this configuration, the charge carriers 
created underneath an interleaved pixel will induce a signal on the read-out 
nodes, capacitively coupled to the interleaved pixel.
In a simplified model, where the sensor is reduced to a capacitive network, 
the ratio of the signal amplitudes on the read-out nodes at the left- 
and right-hand side of the interleaved pixel in both dimensions will be
correlated to the particle position and the resolution is expected to be 
better than ${\rm (implant~pitch)/\sqrt{12}}$ for an implant pitch of 
25~$\mu$m or smaller. The ratio between the 
inter-pixel capacitance and the pixel capacitance to backplane will play a 
crucial role, as it defines the signal amplitude reduction at the output
nodes and therefore the sustainable number of interleaved pixels.
Calculations with such capacitive network models \cite{Pindo} show that 
resolutions similar to those achieved by reading out all pixels are
obtainable if the signal amplitude loss to the backplane is small.
Recent tests on a microstrip sensor, with 200~$\mu$m read-out pitch, 
have achieved a ${\rm 10~\mu m}$ resolution with three interleaved 
strip layout~\cite{Krammer}. Similar results are expected in a pixel
sensor, taking into account both the lower noise because of the
intrinsically smaller load capacitance and the charge sharing
in two dimensions. 
Reducing the read-out density, without compromising the achievable space
resolution, is also beneficial to limit the power dissipation and the 
overall costs. 
In order to verify the feasibility of this scheme, a prototype set of sensors 
with interleaved pixels and different choices of implant and read-out pitch 
have been designed, produced and tested. These test structures have been 
designed in 1998 and delivered in January 1999. Ten high resistivity wafers 
were processed~\footnote{at the Institute of Electron Technology, Warszawa, 
Poland} together with an equal number of low resistivity wafers for process 
control.
A detailed description of the processing can be found in ref.~\cite{AZ}.
A bias grid surrounding the pixel cells allows the polarisation of both the 
interleaved and read-out pixels and each $p^{+}$ pixel implant is connected 
to the metal bias line by polysilicon resistors. This is to ensure a similar
potential for all pixels and hence a uniform charge collection. 
A metal layer is deposited on top of the pixels to be bump-bonded to a 
VLSI cell. The backplane has a meshed metal 
layer to allow the use of an infrared diode for charge collection studies.
Structures with the number of 
interleaved pixels ranging between 0 and 3 were fitted on a 4'' wafer, 
assuming a VLSI cell size of 200 $\times$ 200 or 300 $\times$ 
300~$\mu$m$^{2}$. 
All of the structures on six undiced wafers were visually inspected and
characteristic I--V and C--V curves were measured up to 250~V. The
I--V and 1/C$^2$--V curves obtained for a good structure are shown in 
fig.~\ref{fig:ivcv}. Two wafers have extremely good characteristics, with
a mean current of $\sim$ 50 nA/cm$^2$ and about 50\% of good structures.
The test structures have not shown any design fault even 
if processing and layout optimisation has to be considered. A more
detailed summary of the measurements can be found in ref.~\cite{MCa}.
On a short term, measurements of the inter-pixel and backplane capacitances 
are planned, completing the electrostatics characterisation of these sensors. 
A charge collection study will follow, relying on a low noise analog
chip designed for microstrip sensor read-out and shining an infrared 
light spot on 
the meshed backplane. These measurements will be a proof of principle of 
the proposed device and define the guidelines for further iterations, 
aiming at a 25~$\mu$m pitch.

The device thickness is a particularly relevant issue for the application 
of hybrid pixel sensors at the linear collider. The minimal 
thickness is set by the combination of the sensor noise performance
and the limit of back-thinning technology for a bump-bonded assembly.
Industrial standards guarantee back-thinning down to 50~$\mu$m for the
electronics and procedures to obtain thin sensors are currently being tested 
\cite{SOTT}. The small load capacitance of the pixel cells should
guarantee an extremely good S/N. Scaling what has been obtained for the
microstrip sensors, the desired resolutions should be obtained with a 
200~$\mu$m thick sensor.

\section{Conclusion}

Hybrid pixel sensors are an attractive option for a linear collider Vertex 
Tracker owing to their fast read-out and radiation hardness, well suited to 
the high repetition rate of the TESLA design. The main present limitations of 
this option are the achievable single point space resolution and sensor 
thickness.
To overcome these limitations, pixel sensor design with interleaved cells 
is proposed. Test structures with different configurations of interleaved 
cells have been designed and produced and the results of their first 
electrostatic characterisation are discussed. The conceptual design of a 
Vertex Tracker based on these detectors has been developed and its performances
have been evaluated, using a detailed simulation accounting for the relevant
background processes, and the preliminary results have been presented.

{\bf Acknowledgements}

This activity has been funded in part by the Academy of Finland under the
{\sl R\&D Program for Detectors at Future Colliders} and by MURST under grant 
3418/C.I.


\begin{thebibliography}{999}
\bibitem{Gross} E. Gross, to appear in Proc. of the
{\em Int. Europhysics Conference on High Energy Physics},
Tampere (Finland), July 1999.

\bibitem{higgs1}
H.E. Haber, in Proc. of the {\it  $4^{th}$ Int. Conf. on Physics beyond the 
Standard Model}, Lake Tahoe, (USA); World Scientific, Singapore, 1995;\\
J. Kamoshita, Y. Okada and M. Tanaka, in Proc. of the {\it Workshop on 
Physics and Experiments with Linear Colliders}, Morioka (Japan);
World Scientific, Singapore 1996.

\bibitem{higgs2} M.D. Hildreth, T.L. Barklow and D.L. Burke, 
{\em Phys. Rev. Lett.} {\bf 49} (1994) 3441;\\
M. Battaglia, to appear in Proc. of the {\em Int. Workshop on Linear 
Colliders LCWS99}, Sitges (Spain), May 1999 and hep-ph/9910217.

\bibitem{TopYuk} A. Djouadi, J. Kalinowski and P.M. Zerwas,
{\em Zeit. Phys.} {\bf C 54} (1992) 255.

\bibitem{GMSB} G.F. Giudice and R. Rattazzi, CERN-TH~97-380.

\bibitem{higgs3} G. Borisov, private communication.

\bibitem{lcreview}
K. H\"ubner, in Proc. of the 29$^{th}$ {\em Int. Conf. on High Energy Physics
ICHEP-98}, Vancouver (Canada), July 1998, World Scientific, Singapore, 1999 
and references therein.

\bibitem{CDR} R. Brinkmann, G. Materlik, J. Rossbach, A. Wagner (ed.),
Conceptual Design of 500 GeV e$^-$e$^+$ Linear Collider with Integrated
X-ray Laser Facility, DESY~97-048.

\bibitem{Tesch} N. Tesch, to appear in Proc. of the {\em Int. Workshop
on Linear Colliders LCWS99}, Sitges (Spain), May 1999.

\bibitem{Ye} M. Battaglia and S.~Ye, HIP-1999-73/EXP.

\bibitem{vxd3} K.~Abe {\it et al.},  {\em Nucl. Instr. and Meth.} 
A 400 (1997), 287.

\bibitem{WA97} D. Di Bari {\it et al.}, {\em Nucl. Instr. and Meth.} 
A 395 (1997) 391.

\bibitem{SiTracker} P. Chochula {\it et al.}, {\em Nucl. Instr. and Meth.} 
A 412 (1998) 304.

\bibitem{alice} F. Antinori, in Proc. of the {\em Int. Pixel Detector Workshop
PIXEL98}, FERMILAB-CONF~98-196, 41.

\bibitem{atlas} ATLAS collaboration, The Pixel Detector Technical Design 
Report, CERN-LHCC~98-13. 

\bibitem{cms} D. Bortoletto, in Proc. of the {\em Int. Pixel Detector Workshop
PIXEL98}, FERMILAB-CONF~98-196, 22.

\bibitem{ccdlc} C.J.S. Damerell and D.J. Jackson, in Proc. of the
{\em APS Workshop on New Directions for High Energy Physics}, Snowmass (USA),
July 1996. 

\bibitem{CCDnew} T. Greenshaw {\it et al.}, to appear in Proc. of the
{\em Int. Workshop on Linear Colliders LCWS99}, Sitges (Spain), 
May 1999.

\bibitem{ABC} M. Battaglia, A. Andreazza, M. Caccia and V. Telnov, in Proc. 
of the {\em 2$^{nd}$ Workshop on Backgrounds at Machine-Detector Interface},
Honolulu (USA), March 1997, World Scientific, Singapore, 1998 and in 
DESY~97-123E.

%\bibitem{cmos} E.R. Fossum, {\em IEEE Trans. Electron Devices}
%{\bf 41} (1994) 41.
%\bibitem{pmos} C.J. Kenney {\it et al.}, {\em Nucl. Instr. and Meth.} 
%{\bf A342} (1994) 59.
%\bibitem{cmos2} H. Yamamoto {\it et al.}, in Proc. of the
%{\em Fourth Workshop on Physics and Experiments with Linear Colliders},
%May 1999, Sitges, Spain.

\bibitem{Anna} U. K\"otz {\it et al.}, {\em Nucl. Instr. and Meth.},
A~235 (1985), 481.

\bibitem{Snow} W. Snoeys, CERN--LHCC~97-60, 139.

\bibitem{Bonvicini} V. Bonvicini and M. Pindo, {\em Nucl. Instr. and Meth.}
A~372 (1996) 93.

\bibitem{Pindo} M. Pindo, {\em Nucl. Instr. and Meth.} A~378 (1996) 443.

\bibitem{Krammer} M. Krammer and H. Pernegger, {\em Nucl. Instr. and Meth.}
A~397 (1997) 232.

\bibitem{AZ} W. Kucewicz {\it et al.}, {\em Acta Phys. Pol.}  B~30 (1999) 2075.

\bibitem{MCa} M. Caccia {\it et al.}, to appear in Proc. of the
{\em Int. Workshop on Linear Colliders LCWS99}, Sitges (Spain), 
May 1999 and hep-ex/9910019.

\bibitem{SOTT} T.E. Browder {\it et al.}, in Proc. of the {\em Int. Pixel 
Detector Workshop PIXEL98}, FERMILAB-CONF~98-196, 172.
\end{thebibliography}
\end{document}